\documentclass[twocolumn,prl,aps,superscriptaddress,showpacs,amsmath,amssymb,floatfix]{revtex4-1}
\usepackage{color}
\usepackage{xcolor}
\usepackage{mathrsfs}
\usepackage{amsmath}
\usepackage{graphicx}
\usepackage{dcolumn}
\usepackage{bm}
\usepackage{times}
\usepackage{amssymb}
\usepackage{float}
\usepackage{array}

\newcommand{\bee}{\begin{equation}}
\newcommand{\ene}{\end{equation}}
\newcommand{\bea}{\begin{aligned}}
\newcommand{\ena}{\end{aligned}}

\begin{document}

\title{Majorana Fermions in Semiconducting Nanowires on the Top of Fulde-Ferrell Superconductor}
\author{Jia Liu}
\affiliation{Department of Physics and Center of Coherence, The Chinese University of Hong Kong
Shatin, N.T., Hong Kong, China}%
\author{Chun Fai Chan}
\affiliation{Department of Physics and Center of Coherence, The Chinese University of Hong Kong
Shatin, N.T., Hong Kong, China}%
\author{Ming Gong}
\thanks{skylark.gong@gmail.com}
\affiliation{Department of Physics and Center of Coherence, The Chinese University of Hong Kong
Shatin, N.T., Hong Kong, China}%
\date{\today}

\begin{abstract}
The novel idea that spin-orbit coupling (SOC) and $s$-wave pairing can lead to induced $p$-wave pairing at strong magnetic limit has stimulated widespread interests in the whole community for the searching of Majorana Fermions (MFs), a self-hermitian particle, in semiconductor-superconductor hybrid structures. However, this system has several inherent limitations that prohibit the realization and identification of MFs with the major advances of semiconductor nanotechnology.  We show that these limitations can be resolved by replacing the $s$-wave superconductor with type-II Fulde-Ferrell (FF) superconductor, in which the Cooper pair center-of-mass momentum plays the role of renormalizing the in-plane Zeeman field and chemical potential.  As a result, the MFs can be realized for semiconductor nanowires with small Land\'e $g$ factor and high carrier density. The SOC strength directly influences the topological boundary, thus the topological phase transition and associated MFs can be engineered by an external electric field. Almost all the semiconductor nanowires can be used to realize MFs in this new platform. In particular, we find that InP nanowire, in some aspects, is more suitable for the realization of MFs than InAs and InSb nanowires. This new platform therefore can integrate the advances of semiconductor nanotechnology to the realization and identification of MFs in this hybrid structure.
\end{abstract}


\maketitle

Majorana fermion (MF), a particle which is its own antiparticle\cite{wilczek2009, Majorana}, is the basic building block for fault-tolerant topological quantum computation\cite{qizhangrmp,kanermp}, thus it has been intensively explored in solid materials\cite{readgreen2000,DSarma2005,fukane_2008,fu_2012,kitaev,Jsau,Alicea,Oreg,Lutchyn,Mao, Lee} and ultracold degenerate Fermi gas\cite{zhangcw2008,Sato,gongm2,gongm1,huhui,XJLiu2012,Liang} in the past years. This exotic particle has been predicted more than 80 years for neutrino in particle physics\cite{Majorana}, however, its materialization with quasi-particles is always of great challenge in physics. Breakthrough was made in recent years by the novel idea that spin-orbit coupling (SOC) and $s$-wave pairing can lead to \textit{induced} $p$-wave pairing at strong Zeeman field. This idea can be traced back to the work by Gor’kov and Rashba\cite{Rashba} in 2001 that the SOC (induced by inversion symmetry breaking) can induce mixed singlet  ($s$-wave) and triplet ($p$-wave) pairing in non-centrosymmetric superconductors\cite{NCS1,NCS2,NCS3,NCS4,NCS5,NCS6}. This novel idea has motivated recent theoretical and experimental endeavors in the searching of MFs using semiconductor nanowires on the top of $s$-wave superconductors\cite{Jsau,Alicea,Oreg,Lutchyn,Mao, Lee} (see recent review\cite{RV, RV2}), as well as the realization of topological superfluids in spin-orbit coupled degenerate Fermi gas\cite{gongm2,gongm1,huhui,XJLiu2012}. Recently, some promising signatures\cite{MFexp1,       MFexp2, MFexp3, MFexp4}, though still in heavy debates\cite{debate1, debate2, debate3}, have been reported in a number of experiments based on InAs and InSb nanowires.   These substantial progresses pave a promising way for the realization of MFs and topological quantum computation.

In this scheme, the topological phase can be reached when $V_z^2 > \mu^2 + \Delta^2$\cite{Jsau,Alicea,Oreg,Lutchyn,gongm2,gongm1} , where $V_z$, $\mu$ and $\Delta$ are Zeeman splitting, chemical potential and $s$-wave pairing strength, respectively. Generally, it means that strong magnetic field is required to realize the MFs. There are several basic challenges in experiments to reach this
topological phase. Firstly, the $s$-wave  superconductors generally have very small critical magnetic field ($B_c \sim 1$ Tesla, which corresponds to Zeeman splitting $\sim$ 0.1 meV for Land\'e $g = 2$) \cite{bcmercury, Rohlf}. The Zeeman splitting induced by the critical magnetic field is generally much smaller than $V_z$ unless for nanowires with large Land\'e $g$ factor. Here we do not take the orbital momentum quenching effect into account, which may suppress the $g$ factor due to strong confinement\cite{quenching}. In another word, most of the semiconductor nanowires (with small $g$) are not suitable for the realization of MFs. Secondly, even for InAs and InSb nanowires with large $g$ factor, the chemical potential $|\mu| < |V_z|$ sets another upper bound for carrier density, which is generally very low due to the small effective mass of electron\cite{mingprb}. As a result, the fluctuating effect may become significant, which can destroy the topological phase and associated MFs. Finally, the topological boundary is determined by the Hamiltonian's symmetry at zero momentum (from the viewpoint of Pfaffian\cite{Ghosh}), thus the SOC strength, effective mass and direction of Zeeman field do not directly influence the boundary.
In particular,  the chemical potential in semiconductor is pinned by the Fermi surface of superconductor in the semiconductor-superconductor hybrid structures\cite{mingprb} and can not be tuned by external gate.
Thus the Zeeman field serves as the {\it only} possible parameter to be tuned in experiments\cite{MFexp1, MFexp2, MFexp3, MFexp4}. The above dilemmas are all linked with one another and are unlikely to be solved based on $s$-wave superconductors. So the great advances in semiconductor nanotechnology can not be directly used in this new platform for the searching of MFs.

This paper is aimed to provide a possible solution to the above dilemmas. We show that \textit{all} these limitations can be solved by replacing the $s$-wave superconductor with type-II
Fulde-Ferrell (FF) superconductors, in which the Cooper pairs carry a finite center-of-mass momentum ${\bf Q}$. The basic idea is that: (1) The center-of-mass momentum ${\bf Q}$ plays the role of renormalizing both the in-plane Zeeman field and chemical potential. As a result, the MFs can be realized for semiconductor nanowires with small Land\'e $g$ factor and high carrier density. (2) The SOC strength directly influences the topological boundary, thus the topological phase transition and associated MFs can be tuned by an external electric field, although its chemical potential is still pinned by the Fermi surface of superconductor. (3) Almost {\it all} the Zinc blende and Wurtzite semiconductor nanowires can be used to realize MFs in this new platform. In particular, we find that InP nanowire, in some aspects, is more suitable for the realization of MFs than InAs and InSb nanowires. This new platform therefore can integrate the advances of semiconductor nanotechnology to the realization of MFs in semiconductor-superconductor hybrid structures.

\begin{figure}
\centering
\includegraphics[width=3in]{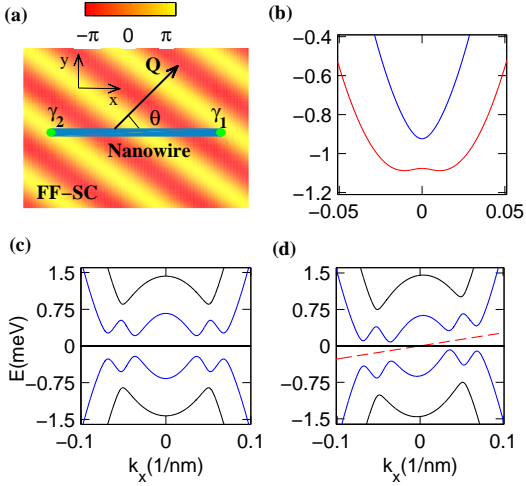}
\caption{(Color online). (a). Semiconductor-FF superconductor hybrid structure for MFs. The phase modulation of the order parameter along direction $\theta$ is sketched by the color bar, and the Cooper pair center-of-mass momentum is denoted by ${\bf Q}$. The Zeeman field is assumed to along $\hat{x}$ direction. (b) Single particle band structure of the InP nanowire using the parameters from Table \ref{tableI}. (c) and (d) are typical band structures for ${\bf Q} = 0$ and ${\bf Q} = 0.01/$nm, respectively. Other parameters are $g\mu B_x = 0.38$ mV and $\theta = \pi/4$. }
\label{fig-fig1}
\end{figure}

Our basic setup is schematically shown in Fig. \ref{fig-fig1}a. The nanowire is in proximity contact with a FF superconductor, which generally has extremely large  critical 
magnetic field ($\sim 10 - 30$ Tesla)\cite{NCS1,NCS2,NCS3,NCS4,NCS5,NCS6}; thus the MFs can still be observed at strong magnetic field. Due to proximity effect, the pairing in nanowire is identical to that in FF superconductors, i.e.,  $\Delta({\bf x}) = \Delta e^{i{\bf Q}\cdot {\bf x}}$. This result has been confirmed by Green's function calculation, see Ref. \onlinecite{chenwei}. So the basic model to describe the hybrid structure reads as ($\hbar = 1$)\cite{Chan,QuNC,WeiZ}
\begin{equation}
\begin{aligned}
&H = H_0+V_{\text{FF}},\\
&H_0 = \frac{{\bf{k}}^2}{2m^*}-\mu+\alpha({\bm{\sigma}}\times{\bf{k}})\cdot\hat{z}+g\mu_B {\bf{B}}\cdot{\bm{\sigma}},\\
&V_{\text{FF}} =  \Delta\sum_{{\bf{k}}}c_{{\bf{k}}+\frac{{\bf{Q}}}{2},\uparrow}^{\dag}c_{-{\bf{k}}+\frac{{\bf{Q}}}{2},\downarrow}^{\dag}+\text{h.c.},
\label{eqH}
\end{aligned}
\end{equation}
where $m^*$ is the effective mass of electron, ${\bf{k}} = (k_x, k_y, k_z)$ are the electron momentum, $\bm{\sigma} = (\sigma_x, \sigma_y, \sigma_z)$ are the
Pauli matrices, ${\bf B} = (B_x, B_y, B_z)$ are the external magnetic fields and $\alpha$ is the Rashba SOC strength.  $c_{{\bf k},s}^\dagger$  ($c_{{\bf k},s}$) is the creation
(anihilation) operator with momentum ${\bf k}$ and spin $s$. The order parameter is set to real without lose of generality.
Eq. \ref{eqH} is obtained via a gauge transformation $c({\bf x}) \rightarrow c({\bf x}) e^{-i{\bf Q}\cdot {\bf x}/2}$ in real space, where ${\bf Q}$ is the corresponding Cooper pair center-of-mass momentum. The corresponding Bogoliubov-de Gennes (BdG) equation reads 
\begin{equation}
H_{\text{BdG}}({\bf{k}})= {{\bf{k}}\cdot{\bf{Q}} \over 2m^*} +  \left(
                   \begin{array}{cc}
                    \bar{H}_{0}({\bf{k}}) & i \Delta\sigma_y \\
                     -i\Delta\sigma_y& -\bar{H}^{*}_{0}(-{\bf{k}}) \\
                   \end{array}
                 \right),
\label{h44m}
\end{equation}
in the Nambu basis $\Psi({\bf{k}})=(c_{{\bf{k}}+{\bf{Q}}/2,\uparrow}, c_{{\bf{k}}+{\bf{Q}}/2,\downarrow}, c_{-{\bf{k}}+{\bf{Q}}/2, \uparrow}^{\dagger}, c_{-{\bf{k}}+{\bf{Q}}/2,\downarrow}^{\dagger})^{T}$. Here $\bar{H}_0({\bf{k}})=\frac{{\bf{k}}^2}{2m^*}-\bar{\mu}+\alpha({\bf{k}}\times{\bm{\sigma}})\cdot\hat{z}+g\mu_B\bf{\bar{B}}\cdot{\bm{\sigma}}$, with $\bar{\mu}=\mu-\frac{|{\bf{Q}|}^2}{8m^*}$, $\bar{B}_x= B_x+\alpha Q_y/2g\mu_B$, $\bar{B}_y = B_y-\alpha Q_x/2g\mu_B$ and $\bar{B}_z=B_z$\cite{Chan}. Define the particle-hole operator as $\Theta=\tau_x K$, where  $\tau_x$ is Pauli matrix acts on particle-hole space and $K$ denotes the complex conjugation, we can verify that $\Theta H_{\text{BdG}}({\bf{k}}) \Theta^{-1} = -H_{\text{BdG}}(-{\bf{k}})$. Thus this system belongs to topological class $D$ with topological index $Z_2$\cite{Ludwig}.

\begin{table}
\caption{Parameters for typical Zinc blende and Wurtzite nanowires used in this work. In the first column the effective mass $m^*$ is in unit of rest electron mass $m_0$, SOC strength $\alpha$ is in unit of  meV$\cdot$nm, $\Delta_{\text{so}} = \alpha^2m^*/2 $ is in unit of $\mu$eV. The parameters for $m^*$, $\alpha$ are from Ref. \onlinecite{material1} and the parameters for Land\'e $g$ factor is from Ref. \onlinecite{material2}.}
\begin{tabular}{|p{0.32in} |p{0.32in} p{0.32in} p{0.32in} p{0.32in} p{0.32in} p{0.32in} |p{0.32in} p{0.32in}|} \hline
&\multicolumn{6}{c|}{Zinc blende}&\multicolumn{2}{c|}{Wurtzite} \\
\cline{2-7} \cline{8-9}
          &InSb  & InAs &GaSb &GaAs  &InP  & Si/Ge & GaN  &AlN        \\ \hline
  $m^*$   & 0.014&0.026 &0.04 &0.063 &0.08 & 0.19& 0.15 &0.25         \\ \hline
 $\alpha$ & 10.0  & 15.0  &10.0    &5.0   &5.0  & 0.06 &0.55 &0.55          \\ \hline
 $\Delta_{\text{so}}$ &9.0 &38.0 &26.0 &10.0 &13.0 &0.004 &0.3 &0.5              \\ \hline
 $g$    & -50.0    & -15.0    & -9.0   & -0.4     & 1.3     & -0.43 & -2.1     & 1.0    \\ \hline
\end{tabular}
\label{tableI}
\end{table}

The decoupling in Eq. \ref{h44m} is exact in free space. Now we explain the physical meaning of Eq. \ref{h44m} in more details. The chemical potential
is balanced out in part by the kinetic energy of Cooper pairs. For typical values ${\bf Q} \sim 0.05 - 0.2$/nm, we can estimate the kinetic energy of Cooper pair in InP nanowire to be about 0.30  - 4.72 meV. This large kinetic energy enables the realization of MFs at relative large carrier density, in which the renormalized $\bar{\mu}$ can still be very small. The center-of-mass momentum also renormalizes the in-plane Zeeman fields. Using typical SOC strength, we can estimate that the Zeeman splitting induced by center-of-mass momentum to be the order of 0.5 - 2.0 meV, which is equivalent to ${\bf B} \sim 10 - 30$ Tesla for typical semiconductor nanowires with small $g$ factors. As a result, the required Zeeman splitting is not necessary to be provided by external Zeeman field and the Land\'e $g$ factor is no longer essential in this new platform. Notice that the SOC, as a standard technique in semiconductor nanotechnology, can be tuned by external electric field. These estimations comprise the key idea of this work. For these reasons, we expect that all the dilemmas mentioned above in $s$-wave superconductor can be resolved in this new platform. In the following we mainly demonstrate our basic idea with InP nanowire\cite{InPNW1, InPNW2}, while in Fig. \ref{fig-fig3} we will summarize the major results using different conventional Zinc blende and Wurtzite semiconductor nanowires; see Table \ref{tableI}.

We first consider a nanowire with strong confinement along its transverse direction, thus $\langle k_{y}\rangle = 0$ and $\langle k_{z}\rangle = 0$.
The contribution of $\langle k_{y}^2\rangle$ and $\langle k_z^2\rangle$ can be absorbed into the chemical potential $\mu$ when only the lowest band along the transverse
direction is occupied, i.e., single-band approximation.  In this case the topological phase of Eq. 1 is determined by Pf$(H_{\text{BdG}} (k = 0)\tau_x)$ = -1,
which yields
\begin{equation}
\bar{\mu}^2 + \Delta^2 < |g\mu_B{\bf B}|^2 +\alpha (g\mu_B{\bf B}\times {\bf Q})\cdot \hat{e}_z+\frac{\alpha^2 |{\bf{Q}}|^2}{4} .
\label{eqB}
\end{equation}
We can recover the well-known result in $s$-wave superconductor, $|g\mu_B{\bf B}|^2 > \mu^2 + \Delta^2$\cite{Jsau,Alicea,Oreg,Lutchyn,gongm2,gongm1}, by setting ${\bf Q} = 0$. Here we see that the boundary is determined not only by the parameters in conventional $s$-wave superconductor ($\mu$, $\Delta$ and $|g\mu_B {\bf B}|$), but also on the direction of external magnetic field, SOC strength and effective mass of Cooper pairs. These new parameters
provide more knobs in experiments for the realization of MFs. In the weak SOC limit, {\bf Q} only renormalizes the 
chemical potential; however, in the strong SOC limit, we have $\bar{\mu}^2 + \Delta^2 < \alpha (g\mu_B{\bf B}\times {\bf Q})\cdot \hat{e}_z+\alpha^2 |{\bf{Q}}|^2/4 \sim \alpha^2 |{\bf Q}|^2/4$, which is independent of Land\'e $g$ factor. Notice that the SOC strength is determined by inversion symmetry, thus can be easily controlled in experiments using external electric field; see Refs. [\onlinecite{SOCE1,SOCE2,SOCE3}]. For this reason, the topological phase transition
can be driven by an external electric field, although the chemical potential is still pinned by the Fermi surface of superconductor\cite{mingprb}.  In other words, the
advances in semiconductor nanotechnology can still be utilized in this hybrid structure to facilitate the realization and identification of MFs.

\begin{figure}
 \center
  \includegraphics[width=3in]{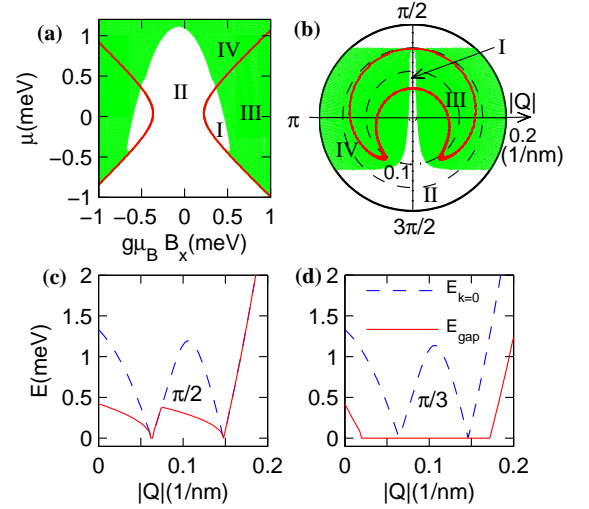}
 \caption{(Color online). (a). Phase diagram as a function of in-plane Zeeman field ${\bf B} = B\hat{x}$ and chemical potential for InP nanowire. The left and right
 parabolas (red lines) are topological
 boundary determined by Eq. 5. The colored and uncolored regimes correspond to gapless phase and gapped phase, respectively. ${\bf Q} = 0.02$/nm and $\theta=\pi/4$ are used.
 (b) Phase diagram in ${\bf Q}$ plane for fixed $\mu = 1$ meV and $g \mu_B B_x = 0.38$ meV. In both figures, I and III are gapped and gapless topological phase, while
 II and IV correspond to trivial gapped and gapless phase, respectively. (c) and (d) plot the energy gap at $k = 0$, and fundamental gap
 $E_{\text{gap}} = \text{min} |E_{n{\bf k}}|$,i.e., the minimal absolute energy gap of the total band structure, as a function of ${\bf Q}$.
 (c) and (d) correspond the result in (b) for $\theta = \pi/2$ and $\theta = \pi/3$.}
\label{fig-fig2}
\end{figure}

To gain a basic insight to this problem, we first present the different phases in Fig. \ref{fig-fig2}a. The interplay beween topology and energy gap gives rise to four
different phases labeled by I to IV\cite{Topologicalprotected}, respectively, where the gapped topological phase in regime I is what needed to search MFs (notice $\theta = \pi/4$ in Fig. \ref{fig-fig2}a). In Fig. \ref{fig-fig2}b, we have fixed all the other parameters but assumed ${\bf Q}$ can rotate in the plane with fixed magnitude. For small ${\bf Q}$ the system is gapped, however, for large ${\bf Q}$ it becomes gapless phase. The most special point in this work is that when ${\bf Q}$ is along $y$ direction (${\bf Q}\perp {\bf B}$), in which condition ${\bf k}\cdot {\bf Q} = 0$ and $({\bf B}\times {\bf Q})\cdot    \hat{e}_z$ becomes maximal  at $\theta = \pi/2$  and minimal at $\theta = 3\pi/2$ (see Fig. \ref{fig-fig2}c). In this case, we can exactly prove that the system is always gapped except at the critical point\cite{DetHQ}. There is a small window when ${\bf Q}$ not exactly perpendicular to the external magnetic field that still supports gapped topological phase (see Figs. \ref{fig-fig2}c-d), and this small window can be controlled in experiments. We have verified that the system is always gapped when $\Delta^2 > \bar{h}_y^2$, thus the small window can be enlarged by 
strong pairing. Moreover, we find that the system can always be gapped when $\bar{h}_x$ (or equivalently $Q_y$) exceeds some critical values, thus we have the stripe-like gapless region in Fig. \ref{fig-fig2}b. Generally, the small regime I can be greatly enlarged by a suitable choice of effective mass as well as SOC strength.

We have also examined the results by taking the multibands along the transverse direction into account. In the simplest case with two bands, the effective
Hamiltonian reads as\cite{multibandRoman}, 
\begin{eqnarray}
H_{\text{BdG}}({\bf{k}}) = && H^{'}_0({\bf k}+\frac{{\bf Q}}{2}) \frac{1+\tau_z}{2} - H^{'*}_0(-{\bf k}+\frac{{\bf Q }}{2})\frac{1-\tau_z}{2} \nonumber \\
&&-\sigma_y(\rho_x|\Delta_{12}|+\Delta_++\rho_z\Delta_-)\tau_y,
\end{eqnarray}
where $H^{'}_0({\bf{k}})= H_0({\bf{k}}) + E_{\text{sp}}\frac{1-\rho_z}{2}-E_{\text{bm}}\sigma_x\rho_y$, with $E_{\text{sp}}=3\pi^2/2mL_y^2$ is the subband energy difference, and $E_{\text{bm}}=8\alpha/3L_y$ is the band mixing energy, which corresponds to the expectation value $\hat{p}_y$
operator between different band eigenstates, $\rho_i$ is the Pauli matrix acts on band degree.  $\Delta_\pm=(\Delta_{11}\pm \Delta_{22})/2$,  where $\Delta_{ij}$ defines pairing strength between band $i$ and $j$. Notice that the system tends to become gapless phase when chemical potential is very large\cite{Chan}, thus only the lowest parabola can support gapped MFs although there are two different parabolas for topological phase. This regime can be realized for a small nanowire with relative large effective mass in experiments. The possible quenching of $g$ factor is not important here\cite{quenching}.

\begin{figure}
\centering
\includegraphics[width=2.4in]{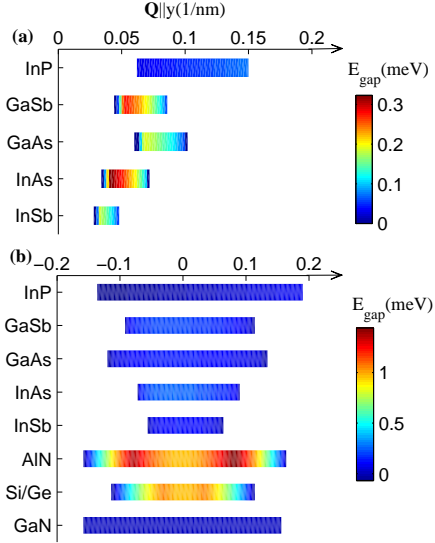}
\caption{(Color online). Possible ${\bf Q} = Q \hat{y}$ for gapped topological phase. (a) $g\mu_B B_x= 0.38$ meV and (b) $g\mu_B B_x= 1.5$ meV. In both figures, $\Delta = 0.3$ meV and $E_{\text{gap}}$ defines the minimal gap of the superconductors (see definition in caption of Fig. \ref{fig-fig2}).}
\label{fig-fig3}
\end{figure}

The basic observations in Fig. \ref{fig-fig2} are quite general. In Fig. \ref{fig-fig3}, we plot the possible ${\bf Q}$ (parallel the $y$ direction) that can support gapped topological phase in different conventional semiconductor nanowires; see Table \ref{tableI}. Fig. \ref{fig-fig3}a are plotted based on initial condition ${\bf Q} = 0$ to be a trivial phase while Fig. \ref{fig-fig3}b is plotted with initial condition to be topological phase. For different nanowires, we use their true effective mass and  SOC as input parameters (see Table \ref{tableI}) for fixed chemical potential and Zeeman splitting $g \mu_B B_x$. These results clearly demonstrate that in almost all the semiconductors, the topological gapped FF phase can always be realized in a wide range of ${\bf Q}$. In particular, we find that InP nanowire which has relative large effective mass and weak SOC strength, in some aspects, is more suitable for the realization of MFs. This result is in stark contrast to the widely accepted belief that the MFs can only be realized in InAs or InSb nanowires due to their large Land\'e $g$ factors\cite{Jsau,Alicea,Oreg,Lutchyn, MFexp1, MFexp2, MFexp3, MFexp4}. In this paper we only consider pure nanowires for simplicity, while in realistic experiments, it is possible to optimize the MFs by considering the alloyed nanowires, in which all the parameters in Table \ref{tableI} can be tuned continually.

\begin{figure}
 \center
 \includegraphics[width=3in]{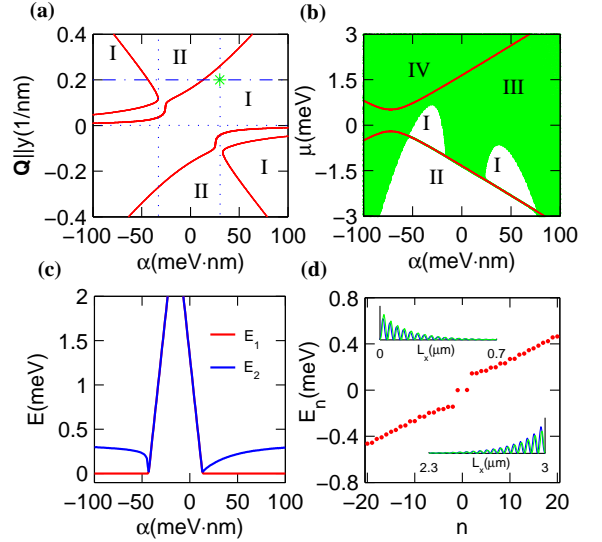}
\caption{(Color online). (a) Phase diagram as a function of ${\bf Q}||y$ and SOC strength for $\mu=1.0$ meV. When $|\alpha|\rightarrow \infty$, the two critical boundaries approach zero according to $Q \sim \frac{2}{\alpha}(-B_x\pm\sqrt{\Delta^2+\mu^2})$, while the other two lines approach infinity according to $Q \sim \pm4m\alpha$. (b) Phase diagram as a function of chemical potential and SOC strength for $|{\bf Q}| = 0.04$/nm, $\theta = 2\pi/5$. (c) and (d) consider a InP nanowire with length $L_x = 3.0$ $\mu$m in contact with a FF superconductor. In (c), we plot the lowest two non-negative eigenvalues $E_1$ and $E_2$ as a function of SOC strength for ${\bf Q}||y = 0.2$/nm (see dash-dotted line in (a)). A typical energy levels at $\alpha = 30$ meV$\cdot$nm (see the green star in (a)) is shown in (d), where the inset show the wave function of MFs. We have verified that $\gamma = \gamma^\dagger$ for the MFs. In all figures, $\Delta = 0.3$ meV and $g\mu_B B_x = 1.5$ meV are used.}
\label{fig-fig4}
\end{figure}

The major advantage of this new platform is that we can utilize the advances of semiconductor nanotechnology to engineer the topological phase transition, although the
chemical potential is still pinned by the Fermi surface of superconductor due to proximity effect\cite{mingprb}. The basic idea relies on the fact that the SOC strength is proportional
to the external electric field\cite{SOCE1, SOCE2, SOCE3}, thus can be tuned in a wide range in realistic experiments. This result opens the possibility to engineer the topological phase transition and associated MFs using an external electric
field instead of magnetic field\cite{MFexp1, MFexp2,MFexp3, MFexp4}. The results are demonstrated in Fig. \ref{fig-fig4}. In Fig. \ref{fig-fig4}a, we show the influence of SOC strength (${\bf Q}||y$) to the gapped trivial phase and gapped topological phase. We show that the SOC strength can dramatically modify the phase diagrams. Especially, the gapped topological phase can be observed in a wide range of ${\bf Q}$ at strong SOC strength (the boundaries are composed by four different ${\bf Q}$ when $\alpha$ exceeds some critical values). Notice that the topological boundary can not intersect with the ${\bf Q} = 0$ line, in which condition the topological boundary is independent of SOC strength. As a result, we find that the two gapped phase at strong SOC are always separated by a trivial phase. The SOC is also possible to drive the system from the gapless phase to the gapped topological phase when ${\bf Q}$ is not along $y$ direction exactly; see Fig. \ref{fig-fig4}b. The results in Figs. \ref{fig-fig4} a-b show that the gapped phase can be dramatically enlarged by controlling the direction of ${\bf Q}$ and the SOC strength.

We also study a realistic system with length $L_x = 3.0$ $\mu$m. We assume the wavefunction to be $\psi = (u_{\uparrow}, u_{\downarrow}, v_{\downarrow}, v_{\uparrow})$,
where $u$ and $v$ are expanded with plane wave basis with a sufficient large cutoff. The basic numerical results are presented in Fig. \ref{fig-fig4}c, in which we see a pair of topological protected MFs (due to $Z_2$ invariant) emerge exactly at the topological phase regime.  A typical energy levels in the topological nanowires are presented in Fig. \ref{fig-fig4}d. The wavefunctions of MFs are well localized at the two ends with exponential decay. The MFs can be denoted as $\gamma = \sum_{s= \uparrow,\downarrow}\int dx u_{s} (x) c_{s} + v_{s} (x) c_{s}^\dagger$, where the self-hermitian, $\gamma = \gamma^\dagger$, requires that $u_{s}=v_{s}^*$, which is also verified in our numerics.

The gapped topological phase depends strongly on the direction of ${\bf Q}$ and the best regime for this phase is ${\bf Q}\perp {\bf B}$. This condition can generally be fulfilled in realistic
experiments. In the type-II FF superconductors, the center-of-mass momentum ${\bf Q}$ is generally produced by an external magnetic field. For a non-centrosymmetric 
superconductor with Rashba SOC, it is well-known that the FF vector ${\bf Q}\perp {\bf B}$\cite{BQ1, BQ2, BQ3, BQ4, Rashba}. This basic conclusion also holds for contact (short range) interaction in ultracold atoms\cite{BQ5, BQ6}. Thus the small topological windows in Fig. \ref{fig-fig2} may be readily realized in this platform without further challenges. Notice that replacing the $s$-wave superconductor with type-II $s$-wave superconductor, which still has large critical field, do not have the above discussed advantages. It is worthwhile to emphasize that the braiding of MFs, in this new platform, can also be achieved by a time-dependent spatial-varying  electric field, which can tune the SOC strength hence the topological phase locally and adiabatically. 

In summary, we introduce a new member to the MF family. We show that the MFs can be realized with almost all conventional Zinc blende and Wurtzite semiconductors with a semiconductor--FF superconductor hybrid structure. In this new platform the topological boundary depends on almost all the parameters of nanowires, thus it provides a lot of knobs for the engineering of topological phase transition. Especially the topological phase transition can be engineered by external electric field. This new platform can integrate the advances of semiconductor nanotechnology to facilitate the realization and identification of MFs in future experiments.

This work is supported by Hong Kong RGC/GRF Projects (No. 401011 and No. 2130352) and  the Chinese   University of Hong Kong (CUHK) Focused Investments Scheme.


\begin{thebibliography}{90}
\bibitem{Topologicalprotected} The topological gapless phase III is still topological protected because it is impossible to adiabatically tune this phase to a trivial phase without closes the energy gap at zero momentum, see Ref. \onlinecite{Chan}.
\bibitem{DetHQ} Det$H_{\text{BdG}}(k_x) = A^2 + \alpha^2 k_x^2$, where $A = \bar{h}_x^2+\bar{h}_y^2-k^4/4 {m^*}^2-\Delta^2-\bar{\mu}^2+k^2(\alpha^2+\mu/m^*)$. Thus the energy gap can close only at the critical boundary; see also discussion in Ref. \onlinecite{gongm2}.
\end{thebibliography}
\end{document}